\documentclass[sigconf]{acmart}

\usepackage{balance}
\usepackage{enumitem}
\usepackage{multirow}

\newcommand{\ignore}[1]{}

\copyrightyear{2017}
\acmYear{2017}
\setcopyright{rightsretained}
\acmConference{SIGIR 2017 Workshop on Neural Information Retrieval (Neu-IR'17)}{}{\\August 7--11, 2017, Shinjuku, Tokyo, Japan} 
\acmPrice{}
\acmDOI{}
\acmISBN{}
\acmPrice{}

\begin{document}

\title[]{Integrating Lexical and Temporal Signals in Neural Ranking Models for Searching Social Media Streams}

\author{Jinfeng Rao,$^{1,3}$ Hua He,$^1$ Haotian Zhang,$^2$ Ferhan Ture,$^3$ \\ Royal Sequiera,$^2$ Salman Mohammed,$^2$ and Jimmy Lin$^2$}
\affiliation{\vspace{0.1cm}
\institution{$^1$ Department of Computer Science, University of Maryland}
\institution{$^2$ David R. Cheriton School of Computer Science, University of Waterloo}
\institution{$^3$ Comcast Applied AI Research Labs}
}

\begin{abstract}
Time is an important relevance signal when searching streams of
social media posts. The distribution of document timestamps from the
results of an initial query can be leveraged to infer the distribution
of relevant documents, which can then be used to rerank the initial
results. Previous experiments have shown that kernel density
estimation is a simple yet effective implementation of this idea.  This
paper explores an alternative approach to mining temporal signals with
recurrent neural networks. Our intuition is that neural networks
provide a more expressive framework to capture the temporal coherence
of neighboring documents in time. To our knowledge, we are
the first to integrate lexical and temporal signals in an end-to-end
neural network architecture, in which existing neural ranking models are used
to generate query--document similarity vectors that feed into a
bidirectional LSTM layer for temporal modeling.
Our results are mixed:\ existing neural models for document ranking alone yield
limited improvements over simple baselines, but the integration of lexical and
temporal signals yield significant improvements over
competitive temporal baselines.
\end{abstract}

\maketitle

\section{Introduction}

There is a large body of literature in information retrieval that has
established the importance of modeling the temporal characteristics
of documents as well as queries for various information
seeking tasks~\cite{LiXiaoyan_Croft_CIKM2003, Efron_Golovchinsky_etal_SIGIR2011,
dakka2012answering, efrontemporal, Elsas_Dumais_WSDM2010, dong2010time, dong2010towards,
Radinsky_etal_WWW2012, radinsky2013behavioral}. Such techniques are particularly important for
searching real-time social media streams such as Twitter, which
rapidly evolves in reaction to real-world events. In this paper,
we tackle the problem of retrospective {\it ad hoc} retrieval over a
collection of short, temporally-ordered social media posts (tweets).
Given information needs expressed as queries, we aim to build 
systems that return high-quality ranked lists of
relevant tweets.

We are motivated by Efron et al.'s temporal cluster
hypothesis~\cite{efrontemporal}, which stipulates that in search
tasks where time plays an important role (such as ours),
relevant documents tend to cluster together in time, and that this
property can be exploited to improve search effectiveness.
Efron et al.~take advantage of kernel density estimation (KDE) to infer
the temporal distribution of relevant documents from an initial search;
the inferred distribution is then used to rerank the original documents. 
Experiments show that this approach is simple yet
effective~\cite{efrontemporal, rao2015reproducible}.

In this paper, we take the KDE technique as a baseline and explore an
alternative approach for temporal modeling using recurrent neural
networks. Such models have been successfully applied to many
sequence learning tasks in natural language processing where the
modeling units are temporally dependent (e.g., tagging and
parsing). We draw a connection between the temporal clustering
of documents, where the relevance of one document may affect its neighbors,
to a sequence learning task, and explore the hypothesis that recurrent neural
networks provide a rich, expressive modeling framework to
capture such temporal signals.
To this end, we build a unified neural network model to
integrate lexical and temporal relevance signals, and we
examine the effectiveness of several existing neural
rankings models that consider only query--document textual similarity. 
We wondered how they would fare
in the context of noisy social media posts.

\smallskip \noindent {\bf Contributions.} We view this work as having
two contributions:
\begin{enumerate}[leftmargin=*]
\item We examined the effectiveness of several existing neural ranking
  models on standard tweet test collections. Results show that, in
  considering only lexical signals, they yield limited improvements
  over simple baselines, suggesting that social media search presents
  a different set of challenges compared to traditional {\it ad hoc} retrieval
  (e.g., over newswire documents).

\item We present, to our knowledge, the first end-to-end neural
  network architecture that integrates lexical and temporal
  signals. Using the best lexical modeling component (from above), we
  are able to obtain significant improvements over competitive
  temporal baselines on standard tweet test collections.

\end{enumerate}

\section{Related Work}

\subsection{Temporal Information Retrieval}

There is a long thread of research exploring the role of temporal
signals in search~\cite{LiXiaoyan_Croft_CIKM2003,
  Efron_Golovchinsky_etal_SIGIR2011, dakka2012answering,
  efrontemporal, Elsas_Dumais_WSDM2010, dong2010time, dong2010towards,
  rao2016timeseries,rao2017termstatistics}, and it is well established
that for certain tasks, better modeling of the temporal
characteristics of queries and documents can lead to higher retrieval
effectiveness.

For example, Jones and Diaz~\cite{JonesRosie_Diaz_TOIS2007} studied
the temporal profiles of queries, classifying queries as atemporal,
temporally ambiguous, or temporally unambiguous. They showed that the
temporal distribution of retrieved documents can provide an additional
source of evidence to improve rankings. Building on this, Li
and Croft~\cite{LiXiaoyan_Croft_CIKM2003} introduced recency priors
that favor more-recent documents. Dakka et
al.~\cite{dakka2012answering} proposed an approach to temporal modeling
based on moving windows to integrate query-specific temporal evidence
with lexical evidence. Efron et
al.~\cite{Efron_Golovchinsky_etal_SIGIR2011} presented several
language modeling variants that incorporate query-specific temporal
evidence. The most direct point of comparison to our work 
(as discussed in the introduction) is the use
of non-parametric density estimation to infer the temporal
distribution of relevant documents from an initial list of retrieved
documents~\cite{efrontemporal, rao2015reproducible}. Most recently,
Rao et al.~\cite{rao2017termstatistics} proposed alternative models
that attempt to make such predictions directly from query term statistics,
obviating the need for an initial retrieval stage.

There have been several other studies of time-based pseudo relevance
feedback. Keikha et al.~\cite{keikha2011temper} represented
queries and documents with their normalized term frequencies in the
time dimension and used a time-based similarity metric to measure
relevance. Craveiro et al.~\cite{craveiro2014query} exploited the
temporal relationship between words for query expansion.
Choi and Croft~\cite{choi2012temporal} presented a method to select time periods
for expansion based on users' behaviors (i.e., retweets). Rao et
al.~\cite{rao2016temporal} proposed a continuous hidden Markov model
to identify temporal burst states in order to select better
query expansion terms.

In addition to ranking, modeling temporal signals has also been shown
to benefit related tasks such as behavior
prediction~\cite{Radinsky_etal_WWW2012, radinsky2013behavioral},
time-sensitive query auto-completion
\cite{Shokouhi_Radinsky_SIGIR2012}, and real-time 
query suggestion~\cite{Mishne_etal_2012}. For example, Radinsky
et al.~\cite{Radinsky_etal_WWW2012, radinsky2013behavioral} built
predictive models to learn query dynamics from historical user data.

\subsection{Neural Information Retrieval}

Following great successes in computer vision, speech recognition, and
natural language processing, we have recently seen a new wave of
research applying neural networks to information retrieval.  Huang et
al.~\cite{huang2013learning} proposed a technique called Deep
Structured Semantic Modeling (DSSM), which has led to follow-on
work~\cite{shen2014latent, song2016multi}. The
basic idea is to use a feedforward function to learn low-dimensional
vector representations of queries and documents, aiming to capture
latent semantic information in texts. Recently, Guo et
al.~\cite{guo2016deep} proposed a deep relevance matching model for
{\it ad hoc} retrieval, pointing out differences between search and
many NLP problems. Mitra et al.~\cite{mitra2017learning} presented a
neural matching model to combine local and global interactions between
queries and documents. There are many other applications of neural networks
to information retrieval, for example, relevance-based word
embeddings~\cite{zamani2017relevance}, voice search with
hierarchical recurrent neural networks~\cite{rao2017talking},
reinforcement learning for query reformulation~\cite{nogueira2017task},
and generative adversarial training for retrieval models~\cite{wang2017irgan}.

On a slightly different thread, there has been work on modeling
textual similarity between short text pairs. Severyn and
Moschitti~\cite{severyn2015learning} proposed a convolutional neural
network (CNN) for exactly this, which was further expanded and
analyzed by Rao et al.~\cite{rao2017cnnqa}. He et
al.~\cite{he2015multi} proposed an ensemble approach of CNNs that take
advantage of different types of convolutional feature maps, pooling
methods, and window sizes to capture sentence pair similarity from
multiple perspectives. Rao et al.~\cite{rao2016noise} extended this
line of work by studying different negative sampling strategies in a
pairwise ranking framework, which obtains state-of-the-art accuracy on
a standard question answering benchmark dataset.

\section{Approach}

\begin{figure}[t]
\center
\includegraphics[width=8.5cm]{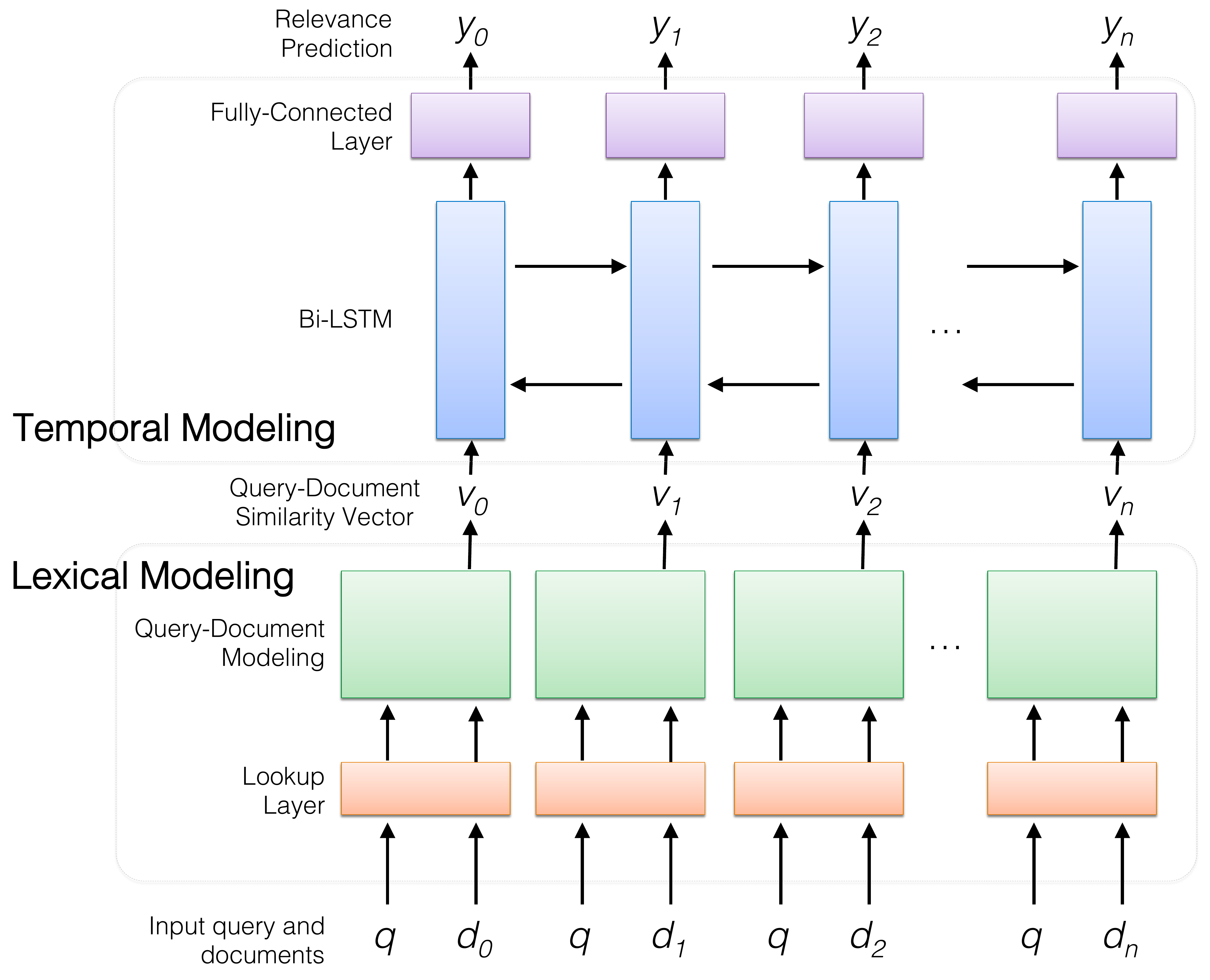}\\
\caption{Our neural network architecture that integrates lexical and
  temporal signals. The lexical modeling component can be viewed as a
  black box for producing query--document similarity vectors. A
  temporally-ordered sequence of these vectors feed into our
  bidirectional LSTM for temporal modeling.}
\label{fig:model}
\end{figure}

We present a neural network architecture that integrates lexical and
temporal signals, shown in Figure~\ref{fig:model}. The overall
architecture consists of distinct components for lexical
modeling, to capture query--document similarity, and temporal modeling, to
capture relevance signals contained in the temporal sequencing of documents.
The two components are independent and in particular we can
view the lexical modeling component as a black box, allowing us to
explore different architectures. However, as we explain later, the
entire model is trained end-to-end in a two-stage process.

\smallskip \noindent {\bf Lexical Modeling.} The architecture for the
lexical modeling component is shown in the lower half of
Figure~\ref{fig:model}, where each ``slice'' of the network is
identical (i.e., with shared parameters). Each instance of the model
takes as input a query and a document to generate a query--document
similarity vector $v$. This is accomplished by translating an input
sequence of tokens (either the query or the document) into a sequence
of distributional vectors $[w_1, w_2, ... w_{|S|}]$, where $|S|$ is
the length of the token sequence, from a word embedding lookup layer. The resulting
matrix then feeds into a neural network. At a high level, this
similarity model can be viewed as a black box, but we describe several
instantiations below.

\smallskip \noindent {\bf Temporal Modeling.} The architecture of the
temporal modeling component is shown in the upper half of
Figure~\ref{fig:model}. We use a bidirectional LSTM where the inputs
are the query--document similarity vectors from the lexical modeling
component, sorted in time order. That is, documents from the training
set are temporally ordered, and the lexical modeling component is
applied to the query paired with each individual document to yield a
collection of query--document similarity vectors $\{v_0, v_1, \ldots,
v_n\}$. The output of the bidirectional LSTM feeds into a
fully-connected layer plus softmax to yield a prediction of document
relevance $y$. Note that each instance of the fully-connected layer and softmax share
parameters.

\smallskip \noindent In what follows, we describe each of the
components in detail.

\subsection{Lexical Modeling Component}
\label{sec:document_embeddings}

In this work, we considered three existing approaches to generating
query--doc\-ument similarity vectors. All three adopt what is
commonly known as a ``Siamese'' structure~\cite{bromley1994signature},
with two subnetworks processing the query and document in parallel,
yielding a ``joined'' representation that feeds into a relevance
modeling component:

\smallskip \noindent \textbf{DSSM}~\cite{huang2013learning}: The Deep
Structured Semantic Model (DSSM) is an early application of neural
networks to web search. One of its key features is a word hashing
layer that converts all tokens into trigrams, which greatly reduces
the size of the vocabulary space to help handle misspellings and
other noisy text input. In parallel, the dense hashed features from
either the query or the document feed into a multi-layer perceptron
with a softmax on top to make the final relevance prediction. We take
the intermediate semantic representation of the query and document,
just before the softmax, as our query--document similarity vector.

\smallskip \noindent \textbf{SM}~\cite{severyn2015learning}: The
convolutional neural network (CNN) proposed by Severyn and
Moschitti~\cite{severyn2015learning} has been previously applied to
question answering as well as tweet reranking. In
both the query and document subnetworks, convolutional feature
maps are applied to the input embedding matrix, followed by ReLU
activation and simple max-pooling, to arrive at a representation
vector $\textrm{x}_q$ for the query and $\textrm{x}_d$ for the
document. Intermediate representations are concatenated
into a single vector at the join layer:
\begin{equation}
\textrm{x}_{\textrm{join}} = [ \textrm{x}_q^T; x_{\textrm{sim}} ; \textrm{x}_d^T; \textrm{x}_{\textrm{feat}}^T ]
\end{equation}
\noindent where $x_{\textrm{sim}}$ defines the \emph{bilinear} similarity between
$\textrm{x}_q$ and $\textrm{x}_d$.
The final component consists of
``extra features'' $\textrm{x}_{\textrm{feat}}$ derived from four word
overlap measures between the query and the document.

In the original SM model, the join vector feeds into a fully-connected
layer and softmax for final relevance prediction, but in our
approach we use the join vector $\textrm{x}_{\textrm{join}}$ as the
query--document similarity vector.

\smallskip \noindent \textbf{Multi-Perspective
  CNN}~\cite{he2015multi}: This approach was developed at roughly the
same time as the SM model and can be described as an ensemble of
convolutional neural networks. The ``multi-perspective'' idea refers
to different types of convolutional feature maps, pooling methods, and
window sizes to capture semantic similarity between textual
inputs. Another key feature is a similarity measurement layer to
explore the interactions between the learned convolutional
feature maps at different levels of granularity. At the time the work
was published, it achieved state-of-the-art effectiveness on several
semantic modeling tasks such as paraphrase detection and question
answering (although other models have improved upon it since).

As with the SM model, we take the joined representation just before
the fully-connected layer and softmax as the query--document
similarity vector.

\subsection{Temporal Modeling Component}
\label{sec:temporal_modeling}

On top of a sequence of temporally ordered query--document similarity
vectors (the output of the lexical modeling component), we layer a
recurrent neural network to capture the temporal clustering of
relevant documents (see Figure~\ref{fig:model}). Compared to kernel
density estimation, we hypothesized that recurrent neural networks
provide a richer, more expressive modeling framework to capture
temporal signals that can yield more effective results.

For our task, we used a variant of recurrent neural networks,
bidirectional LSTM
(Bi-LSTM)~\cite{Hochreiter:1997:LSM:1246443.1246450}, which have been
successfully applied to text similarity
tasks~\cite{He2016PairwiseWI,DBLP:conf/semeval/HeWGRL16}. One key
feature of LSTMs is their ability to capture long-range dependencies,
and a bidirectional LSTM consists of two LSTMs that run in parallel in
opposite directions:\ one (forward $\textrm{LSTM}^{f}$) on the input sequence
and the other (backward $\textrm{LSTM}^{b}$) on the reverse of the sequence. At
time step $t$, the Bi-LSTM hidden state $h^{\textrm{bi}}_{t}$ is a concatenation
of the hidden state $h^{\textrm{for}}_t$ of $\textrm{LSTM}^{f}$ and the hidden state
$h^{\textrm{back}}_t$ of $\textrm{LSTM}^{b}$, representing the neighboring contexts of
input $v_t$ in the temporal sequence.

Given Bi-LSTM output $h^{\textrm{bi}}_{t}$, the prediction output $y_t$
of our temporal ranking model at time step $t$ is obtained by passing
the Bi-LSTM output through a fully-connected layer and softmax as
follows:
\begin{gather}\label{header}
g_t = \sigma(W^{m}\cdot h^{\textrm{bi}}_{t} + b^{m})\\
y_t = \textrm{softmax}(W^{p}\cdot g_t + b^{p})
\end{gather}
\noindent where the output $y_t$ indicates the relevance of the
document at time step $t$. $W^*$ and $b^*$ are learned
weight matrices and biases.

\subsection{Model Training}

Although our neural network architecture breaks down into two distinct
components, we train the entire model end-to-end in a two-stage
manner, with stochastic gradient descent to minimize negative
log-likelihood loss of the entire model. In each epoch, we first train
the lexical modeling component independently, and then use the results
to generate inputs to the temporal modeling layer. The losses from all
documents are summed together to train the Bi-LSTM and the top layers, while the
underlying lexical component is held constant. The reason for this
two-stage approach is to restrict the search space during model
optimization, since we have limited labeled data for training.

At inference time, we first retrieve candidate documents from the
collection using a standard ranking function. These documents are then
ordered chronologically and fed into the model. The classification
scores outputted by each step of the Bi-LSTM (corresponding to the
processing of that document) are used to resort the ranked list, which
we take as final output for evaluation.

\begin{table*}[t]
\centering
\begin{tabular}{| l| ll| l| l| l| l|}
\hline
{\bf ID} &\multicolumn{2}{c|}{\bf Method} & {\bf P15} & {\bf P30} & {\bf P100} & {\bf AP} \\
\hline
\hline
1 &\multicolumn{2}{l|}{Query Likelihood (QL)~\cite{Ponte98}} & 0.381 & 0.329 & 0.234 & 0.200  \\
\hline
\hline
\multicolumn{7}{|c|}{\textbf{Temporal Baselines}}\\ \hline 
2 & \multirow{4}{*}{KDE~\cite{efrontemporal}} & uniform & 0.366 & 0.326 & 0.243 & 0.203  \\
3 &									& score-based & 0.383 & 0.334 & 0.244 & 0.203 \\
4 &									& rank-based  & 0.387 & 0.337 & 0.244 & 0.204 \\
5 &                                                                       & oracle & \textbf{0.411$^{1,4}$} & \textbf{0.389$^{1,4}$} & \textbf{0.260$^{1,4}$} & \textbf{0.229$^{1,4}$} \\
\hline
\hline
\multicolumn{7}{|c|}{\textbf{Neural Ranking Approaches}}\\ \hline 
6 &\multicolumn{2}{l|}{DSSM~\cite{huang2013learning}} & 0.187 & 0.168 & 0.153 & 0.102  \\ \hline
7 & \multicolumn{2}{l|}{SM~\cite{severyn2015learning}} & 0.203 & 0.188 & 0.170 & 0.116  \\ \hline
8 &\multicolumn{2}{l|}{Multi-Perspective CNN~\cite{he2015multi}} & \textbf{0.401$^{1}$} & \textbf{0.356$^{1}$} & \textbf{0.252$^{1}$} & \textbf{0.197}  \\ \hline
\hline
\multicolumn{7}{|c|}{\textbf{Neural Ranking + Temporal Modeling}} \\ \hline 
9 & \multicolumn{2}{l|}{SM~\cite{severyn2015learning} + Temporal} & 0.222 & 0.196 & 0.169 & 0.116  \\ \hline
10 &\multicolumn{2}{l|}{Multi-Perspective CNN~\cite{he2015multi} + Temporal} & \textbf{0.418$^{1,4,8}$} & \textbf{0.366$^{1,4}$} & \textbf{0.257$^{1,4,8}$} & \textbf{0.203}$^{1,8}$  \\ \hline
\end{tabular}
\vspace{0.25cm}
\caption{Results from the TREC 2011/2012 Microblog Track test collections,
  using TREC 2011 data for training and TREC 2012 data for evaluation.
  Superscripts indicate the row indexes from which the metric
  difference is statistically significant ($p < 0.05$) using Fisher's
  two-sided, paired randomization test.}
\label{table:results-2011-2012}
\vspace{-0.3cm}
\end{table*}

\section{Experiments}

\subsection{Experimental Setup}

We evaluated our proposed models on Twitter test collections from the
TREC 2011 and 2012 Microblog Tracks (49 topics and 59 topics,
respectively). Both use the Tweets2011 collection, which consists of
an approximately 1\% sample (after some spam removal) of tweets from
January 23, 2011 to February 7, 2011 (inclusive), totaling
approximately 16 million tweets. Relevance judgments were made on a
3-point scale (``not relevant'', ``relevant'', ``highly relevant''),
but in this work we treated both higher grades as relevant. We removed
all the retweets in our experiments since they are by definition not
relevant according to the assessment guidelines.

To rule out the effects of different preprocessing strategies during
collection preparation (i.e., stemming, stopword removal, etc.), we
used the open-source implementations of tweet search provided by the
TREC Microblog
API\footnote{\url{https://github.com/lintool/twitter-tools}} to
retrieve up to 1000 tweets per topic using query likelihood (QL) for
scoring. These initial results were then reranked using our proposed
models. For effectiveness, we measured average precision (AP) and
precision at 15, 30, and 100 (P15, P30, and P100); note that P30 was the
official metric used in the TREC Microblog Tracks. Since all the
models required training, we used the TREC 2011 topics for training
and the TREC 2012 topics for evaluation. Additionally, we randomly
selected 5\% of query-document pairs from the training set as the
development set; those selected samples were removed from the training
set.

We considered several lexical and temporal baselines to evaluate our
models. The standard query likelihood (QL) approach~\cite{Ponte98} was
used as the lexical baseline. We used the kernel density estimation
techniques of Efron et al.~\cite{efrontemporal} as our temporal
baseline (with the implementation from Rao et al.~\cite{rao2015reproducible}).
They proposed four different weighting schemes to estimate
feedback parameters:\ uniform, score-based, rank-based, and
oracle. The first three take advantage of document timestamp
distributions from an initial retrieval, while the oracle method
requires actual human relevance judgments. The oracle is useful to
illustrate upper bound effectiveness for KDE-based techniques.

The neural ranking approaches were implemented using the Torch deep
learning toolkit (in Lua). For the SM
model\footnote{https://github.com/castorini/SM-CNN-Torch}~\cite{severyn2015learning}
and the multi-perspective CNN\footnote{https://github.com/castorini/MP-CNN-Torch}~\cite{he2015multi},
we took advantage of existing open-source implementations; DSSM is our
own re-implementation. We used existing 300-dimensional
GloVe~\cite{pennington2014glove} word embeddings to encode each word,
which was trained on 840 billion tokens and freely available. The
vocabulary size of our dataset is 90.3K, with around 37\% words not found
in the GloVe word embeddings. Unknown words were randomly initialized
with values uniformly sampled from [$-$0.05, 0.05]. During training, we
used stochastic gradient descent together with RMS-PROP to iteratively
update the model. The output size of the Bi-LSTM layer is 400 and the
hidden layer size is 150.  The learning rate was initially set to
$0.001$, and then decreased by a factor of three when the development
set loss stopped decreasing for three epochs. The maximum number of
training epochs was 25.

\subsection{Experimental Results}

Table~\ref{table:results-2011-2012} shows our experimental results,
with each row representing an experimental condition (numbered for
convenience). For each method, we performed significance testing
against the lexical baseline (QL) and the best-performing temporal KDE
model (rank-based). In addition, we tested the significance of
differences between each pair of lexical-only model vs.\ lexical +
temporal model. In all cases, we used Fisher's two-sided, paired
randomization test~\cite{smucker2007comparison}. Superscripts indicate
the row indexes for which the metric difference is statistically
significant ($p < 0.05$).

From the block in Table~\ref{table:results-2011-2012} labeled
``Temporal Baselines'', we see that the KDE approaches (with the
exception of the oracle condition) yield limited improvements over the
QL baseline.\footnote{These results are consistent with those of Rao
  et al.~\cite{rao2015reproducible}; although those experiments affirmed
  the overall effectiveness of the KDE techniques, results from
  individual configurations (such as a particular train/test split)
  may not yield significant improvements.} Looking at the block of
Table~\ref{table:results-2011-2012} labeled ``Neural Ranking
Approaches'', we find that the SM model and DSSM do not appear to be
as effective as the multi-perspective CNN; in particular, the first two
models actually perform worse than the simple QL baseline.

In Table~\ref{table:results-2011-2012}, under ``Neural Ranking +
Temporal Modeling'', we report results from combining the SM model and
the multi-perspective CNN with our Bi-LSTM temporal model. In the
first case, the improvement is minor over the SM model alone, but with
the multi-perspective CNN, the addition of a temporal layer yields
significant improvements over the multi-perspective CNN alone
(condition 8) and also rank-based KDE (condition 4). We also note that
the multi-perspective CNN + Bi-LSTM model approaches the effectiveness
of the oracle KDE condition (and in the case of P15, exceeds it,
albeit not significantly). This suggests that neural networks offer an
expressive framework for integrating lexical and temporal signals,
potentially beyond what is available to non-parametric density
estimation techniques alone, even with oracle input.

\section{Future Work and Conclusions}

While our results are certainly encouraging, there are a number of
unresolved issues and open questions; these are avenues for future
work. First, we have only experimented on a single collection of
tweets, and thus there are questions about the robustness of our
results. Second, we have yet to perform detailed error analysis to
uncover the differences between the three neural ranking models we
examined, and thus have not answered the {\it why} questions:\ For
example, what characteristics of the multi-perspective CNN allow it to
serve as an effective ranker while the SM model and DSSM do not appear
to work? As a start, a topic-by-topic analysis might uncover
some insights. Third, it is interesting to note that our approaches
improve early precision more than average precision:\ it is unclear if
this is due to inherent properties of our model, our reranking setup,
or some other reason.

To conclude, we believe that this work is most valuable in providing a
general architecture for integrating lexical and temporal signals for
information seeking on time-ordered documents. We have already shown
that different lexical modeling components can be ``plugged in''---our
experiments examined three neural network models, but more can be
straightforwardly explored. Similarly, we can imagine different
temporal models beyond the Bi-LSTM approach proposed here. In
addition, we have shown that our combined lexical and temporal models
can be trained end to end, which yields an integrated, flexible, and
expressive ranking framework.

\section{Acknowledgments} 

This work was supported by the Natural Sciences and Engineering
Research Council (NSERC) of Canada, with additional contributions from
the U.S.\ National Science Foundation under CNS-1405688. Any findings,
conclusions, or recommendations express\-ed do not necessarily reflect
the views of the sponsors.

\balance
%\bibliographystyle{ACM-Reference-Format}
%\bibliography{nn-temporal-ranking}

\begin{thebibliography}{00}

%%% the Web address.

\ifx \showCODEN    \undefined \def \showCODEN     #1{\unskip}     \fi
\ifx \showDOI      \undefined \def \showDOI       #1{{\tt DOI:}\penalty0{#1}\ }
  \fi
\ifx \showISBNx    \undefined \def \showISBNx     #1{\unskip}     \fi
\ifx \showISBNxiii \undefined \def \showISBNxiii  #1{\unskip}     \fi
\ifx \showISSN     \undefined \def \showISSN      #1{\unskip}     \fi
\ifx \showLCCN     \undefined \def \showLCCN      #1{\unskip}     \fi
\ifx \shownote     \undefined \def \shownote      #1{#1}          \fi
\ifx \showarticletitle \undefined \def \showarticletitle #1{#1}   \fi
\ifx \showURL      \undefined \def \showURL       #1{#1}          \fi
% invisible to TeX
\providecommand\bibfield[2]{#2}
\providecommand\bibinfo[2]{#2}
\providecommand\natexlab[1]{#1}
\providecommand\showeprint[2][]{arXiv:#2}

\bibitem[\protect\citeauthoryear{Bromley, Guyon, LeCun, S\"{a}ckinger, and
  Shah}{Bromley et~al\mbox{.}}{1993}]%
        {bromley1994signature}
\bibfield{author}{\bibinfo{person}{Jane Bromley}, \bibinfo{person}{Isabelle
  Guyon}, \bibinfo{person}{Yann LeCun}, \bibinfo{person}{Eduard S\"{a}ckinger},
  {and} \bibinfo{person}{Roopak Shah}.} \bibinfo{year}{1993}\natexlab{}.
\newblock \showarticletitle{Signature Verification Using a ``Siamese'' Time
  Delay Neural Network}. In \bibinfo{booktitle}{{\em NIPS}}.
  \bibinfo{pages}{737--744}.
\newblock


\bibitem[\protect\citeauthoryear{Choi and Croft}{Choi and Croft}{2012}]%
        {choi2012temporal}
\bibfield{author}{\bibinfo{person}{Jaeho Choi} {and} \bibinfo{person}{W.~Bruce
  Croft}.} \bibinfo{year}{2012}\natexlab{}.
\newblock \showarticletitle{Temporal Models for Microblogs}. In
  \bibinfo{booktitle}{{\em CIKM}}. \bibinfo{pages}{2491--2494}.
\newblock


\bibitem[\protect\citeauthoryear{Craveiro, Macedo, and Madeira}{Craveiro
  et~al\mbox{.}}{2014}]%
        {craveiro2014query}
\bibfield{author}{\bibinfo{person}{Olga Craveiro}, \bibinfo{person}{Joaquim
  Macedo}, {and} \bibinfo{person}{Henrique Madeira}.}
  \bibinfo{year}{2014}\natexlab{}.
\newblock \showarticletitle{Query Expansion with Temporal Segmented Texts}. In
  \bibinfo{booktitle}{{\em ECIR}}. \bibinfo{pages}{612--617}.
\newblock


\bibitem[\protect\citeauthoryear{Dakka, Gravano, and Ipeirotis}{Dakka
  et~al\mbox{.}}{2012}]%
        {dakka2012answering}
\bibfield{author}{\bibinfo{person}{Wisam Dakka}, \bibinfo{person}{Luis
  Gravano}, {and} \bibinfo{person}{Panagiotis~G. Ipeirotis}.}
  \bibinfo{year}{2012}\natexlab{}.
\newblock \showarticletitle{Answering General Time-Sensitive Queries}.
\newblock \bibinfo{journal}{{\em TKDE\/}} \bibinfo{volume}{24},
  \bibinfo{number}{2} (\bibinfo{year}{2012}), \bibinfo{pages}{220--235}.
\newblock


\bibitem[\protect\citeauthoryear{Dong, Chang, Zheng, Mishne, Bai, Zhang,
  Buchner, Liao, and Diaz}{Dong et~al\mbox{.}}{2010a}]%
        {dong2010towards}
\bibfield{author}{\bibinfo{person}{Anlei Dong}, \bibinfo{person}{Yi Chang},
  \bibinfo{person}{Zhaohui Zheng}, \bibinfo{person}{Gilad Mishne},
  \bibinfo{person}{Jing Bai}, \bibinfo{person}{Ruiqiang Zhang},
  \bibinfo{person}{Karolina Buchner}, \bibinfo{person}{Ciya Liao}, {and}
  \bibinfo{person}{Fernando Diaz}.} \bibinfo{year}{2010}\natexlab{a}.
\newblock \showarticletitle{Towards Recency Ranking in Web Search}. In
  \bibinfo{booktitle}{{\em WSDM}}. \bibinfo{pages}{11--20}.
\newblock


\bibitem[\protect\citeauthoryear{Dong, Zhang, Kolari, Bai, Diaz, Chang, Zheng,
  and Zha}{Dong et~al\mbox{.}}{2010b}]%
        {dong2010time}
\bibfield{author}{\bibinfo{person}{Anlei Dong}, \bibinfo{person}{Ruiqiang
  Zhang}, \bibinfo{person}{Pranam Kolari}, \bibinfo{person}{Jing Bai},
  \bibinfo{person}{Fernando Diaz}, \bibinfo{person}{Yi Chang},
  \bibinfo{person}{Zhaohui Zheng}, {and} \bibinfo{person}{Hongyuan Zha}.}
  \bibinfo{year}{2010}\natexlab{b}.
\newblock \showarticletitle{Time is of the Essence: Improving Recency Ranking
  Using Twitter Data}. In \bibinfo{booktitle}{{\em WWW}}.
  \bibinfo{pages}{331--340}.
\newblock


\bibitem[\protect\citeauthoryear{Efron and Golovchinsky}{Efron and
  Golovchinsky}{2011}]%
        {Efron_Golovchinsky_etal_SIGIR2011}
\bibfield{author}{\bibinfo{person}{Miles Efron} {and} \bibinfo{person}{Gene
  Golovchinsky}.} \bibinfo{year}{2011}\natexlab{}.
\newblock \showarticletitle{Estimation Methods for Ranking Recent Information}.
  In \bibinfo{booktitle}{{\em SIGIR}}. \bibinfo{pages}{495--504}.
\newblock


\bibitem[\protect\citeauthoryear{Efron, Lin, He, and de~Vries}{Efron
  et~al\mbox{.}}{2014}]%
        {efrontemporal}
\bibfield{author}{\bibinfo{person}{Miles Efron}, \bibinfo{person}{Jimmy Lin},
  \bibinfo{person}{Jiyin He}, {and} \bibinfo{person}{Arjen de Vries}.}
  \bibinfo{year}{2014}\natexlab{}.
\newblock \showarticletitle{Temporal Feedback for Tweet Search with
  Non-Parametric Density Estimation}. In \bibinfo{booktitle}{{\em SIGIR}}.
  \bibinfo{pages}{33--42}.
\newblock


\bibitem[\protect\citeauthoryear{Elsas and Dumais}{Elsas and Dumais}{2010}]%
        {Elsas_Dumais_WSDM2010}
\bibfield{author}{\bibinfo{person}{Jonathan~L. Elsas} {and}
  \bibinfo{person}{Susan~T. Dumais}.} \bibinfo{year}{2010}\natexlab{}.
\newblock \showarticletitle{Leveraging Temporal Dynamics of Document Content in
  Relevance Ranking}. In \bibinfo{booktitle}{{\em WSDM}}.
  \bibinfo{pages}{1--10}.
\newblock


\bibitem[\protect\citeauthoryear{Guo, Fan, Ai, and Croft}{Guo
  et~al\mbox{.}}{2016}]%
        {guo2016deep}
\bibfield{author}{\bibinfo{person}{Jiafeng Guo}, \bibinfo{person}{Yixing Fan},
  \bibinfo{person}{Qingyao Ai}, {and} \bibinfo{person}{W.~Bruce Croft}.}
  \bibinfo{year}{2016}\natexlab{}.
\newblock \showarticletitle{A Deep Relevance Matching Model for Ad-hoc
  Retrieval}. In \bibinfo{booktitle}{{\em CIKM}}. \bibinfo{pages}{55--64}.
\newblock


\bibitem[\protect\citeauthoryear{He, Gimpel, and Lin}{He et~al\mbox{.}}{2015}]%
        {he2015multi}
\bibfield{author}{\bibinfo{person}{Hua He}, \bibinfo{person}{Kevin Gimpel},
  {and} \bibinfo{person}{Jimmy Lin}.} \bibinfo{year}{2015}\natexlab{}.
\newblock \showarticletitle{Multi-Perspective Sentence Similarity Modeling with
  Convolutional Neural Networks}. In \bibinfo{booktitle}{{\em EMNLP}}.
  \bibinfo{pages}{1576--1586}.
\newblock


\bibitem[\protect\citeauthoryear{He and Lin}{He and Lin}{2016}]%
        {He2016PairwiseWI}
\bibfield{author}{\bibinfo{person}{Hua He} {and} \bibinfo{person}{Jimmy Lin}.}
  \bibinfo{year}{2016}\natexlab{}.
\newblock \showarticletitle{Pairwise Word Interaction Modeling with Deep Neural
  Networks for Semantic Similarity Measurement}. In \bibinfo{booktitle}{{\em
  HLT-NAACL}}. \bibinfo{pages}{937--948}.
\newblock


\bibitem[\protect\citeauthoryear{He, Wieting, Gimpel, Rao, and Lin}{He
  et~al\mbox{.}}{2016}]%
        {DBLP:conf/semeval/HeWGRL16}
\bibfield{author}{\bibinfo{person}{Hua He}, \bibinfo{person}{John Wieting},
  \bibinfo{person}{Kevin Gimpel}, \bibinfo{person}{Jinfeng Rao}, {and}
  \bibinfo{person}{Jimmy Lin}.} \bibinfo{year}{2016}\natexlab{}.
\newblock \showarticletitle{{UMD-TTIC-UW} at SemEval-2016 Task 1:
  Attention-Based Multi-Perspective Convolutional Neural Networks for Textual
  Similarity Measurement}. In \bibinfo{booktitle}{{\em SemEval}}.
  \bibinfo{pages}{1103--1108}.
\newblock


\bibitem[\protect\citeauthoryear{Hochreiter and Schmidhuber}{Hochreiter and
  Schmidhuber}{1997}]%
        {Hochreiter:1997:LSM:1246443.1246450}
\bibfield{author}{\bibinfo{person}{Sepp Hochreiter} {and}
  \bibinfo{person}{J{\"u}rgen Schmidhuber}.} \bibinfo{year}{1997}\natexlab{}.
\newblock \showarticletitle{Long Short-Term Memory}.
\newblock \bibinfo{journal}{{\em Neural Computation\/}} \bibinfo{volume}{9},
  \bibinfo{number}{8} (\bibinfo{year}{1997}), \bibinfo{pages}{1735--1780}.
\newblock


\bibitem[\protect\citeauthoryear{Huang, He, Gao, Deng, Acero, and Heck}{Huang
  et~al\mbox{.}}{2013}]%
        {huang2013learning}
\bibfield{author}{\bibinfo{person}{Po-Sen Huang}, \bibinfo{person}{Xiaodong
  He}, \bibinfo{person}{Jianfeng Gao}, \bibinfo{person}{Li Deng},
  \bibinfo{person}{Alex Acero}, {and} \bibinfo{person}{Larry Heck}.}
  \bibinfo{year}{2013}\natexlab{}.
\newblock \showarticletitle{Learning Deep Structured Semantic Models for Web
  Search using Clickthrough Data}. In \bibinfo{booktitle}{{\em CIKM}}.
  \bibinfo{pages}{2333--2338}.
\newblock


\bibitem[\protect\citeauthoryear{Jones and Diaz}{Jones and Diaz}{2007}]%
        {JonesRosie_Diaz_TOIS2007}
\bibfield{author}{\bibinfo{person}{Rosie Jones} {and} \bibinfo{person}{Fernando
  Diaz}.} \bibinfo{year}{2007}\natexlab{}.
\newblock \showarticletitle{Temporal Profiles of Queries}.
\newblock \bibinfo{journal}{{\em TOIS\/}}
  \bibinfo{volume}{25}, \bibinfo{number}{3} (\bibinfo{year}{2007}),
  \bibinfo{pages}{Article 14}.
\newblock


\bibitem[\protect\citeauthoryear{Keikha, Gerani, and Crestani}{Keikha
  et~al\mbox{.}}{2011}]%
        {keikha2011temper}
\bibfield{author}{\bibinfo{person}{Mostafa Keikha}, \bibinfo{person}{Shima
  Gerani}, {and} \bibinfo{person}{Fabio Crestani}.}
  \bibinfo{year}{2011}\natexlab{}.
\newblock \showarticletitle{TEMPER: A Temporal Relevance Feedback Method}. In
  \bibinfo{booktitle}{{\em ECIR}}. \bibinfo{pages}{436--447}.
\newblock


\bibitem[\protect\citeauthoryear{Li and Croft}{Li and Croft}{2003}]%
        {LiXiaoyan_Croft_CIKM2003}
\bibfield{author}{\bibinfo{person}{Xiaoyan Li} {and} \bibinfo{person}{W.~Bruce
  Croft}.} \bibinfo{year}{2003}\natexlab{}.
\newblock \showarticletitle{Time-Based Language Models}. In
  \bibinfo{booktitle}{{\em CIKM}}. \bibinfo{pages}{469--475}.
\newblock


\bibitem[\protect\citeauthoryear{Mishne, Dalton, Li, Sharma, and Lin}{Mishne
  et~al\mbox{.}}{2012}]%
        {Mishne_etal_2012}
\bibfield{author}{\bibinfo{person}{Gilad Mishne}, \bibinfo{person}{Jeff
  Dalton}, \bibinfo{person}{Zhenghua Li}, \bibinfo{person}{Aneesh Sharma},
  {and} \bibinfo{person}{Jimmy Lin}.} \bibinfo{year}{2012}\natexlab{}.
\newblock \showarticletitle{Fast Data in the Era of Big Data: {Twitter's}
  Real-Time Related Query Suggestion Architecture}. In \bibinfo{booktitle}{{\em
  SIGMOD}}. \bibinfo{pages}{1147--1157}.
\newblock


\bibitem[\protect\citeauthoryear{Mitra, Diaz, and Craswell}{Mitra
  et~al\mbox{.}}{2017}]%
        {mitra2017learning}
\bibfield{author}{\bibinfo{person}{Bhaskar Mitra}, \bibinfo{person}{Fernando
  Diaz}, {and} \bibinfo{person}{Nick Craswell}.}
  \bibinfo{year}{2017}\natexlab{}.
\newblock \showarticletitle{Learning to Match using Local and Distributed
  Representations of Text for Web Search}. In \bibinfo{booktitle}{{\em WWW}}.
  \bibinfo{pages}{1291--1299}.
\newblock


\bibitem[\protect\citeauthoryear{Nogueira and Cho}{Nogueira and Cho}{2017}]%
        {nogueira2017task}
\bibfield{author}{\bibinfo{person}{Rodrigo Nogueira} {and}
  \bibinfo{person}{Kyunghyun Cho}.} \bibinfo{year}{2017}\natexlab{}.
\newblock \showarticletitle{Task-Oriented Query Reformulation with
  Reinforcement Learning}.
\newblock \bibinfo{journal}{{\em arXiv:1704.04572\/}}.
\newblock


\bibitem[\protect\citeauthoryear{Pennington, Socher, and Manning}{Pennington
  et~al\mbox{.}}{2014}]%
        {pennington2014glove}
\bibfield{author}{\bibinfo{person}{Jeffrey Pennington},
  \bibinfo{person}{Richard Socher}, {and} \bibinfo{person}{Christopher~D.
  Manning}.} \bibinfo{year}{2014}\natexlab{}.
\newblock \showarticletitle{Glove: Global Vectors for Word Representation}. In
  \bibinfo{booktitle}{{\em EMNLP}}. \bibinfo{pages}{1532--1543}.
\newblock


\bibitem[\protect\citeauthoryear{Ponte and Croft}{Ponte and Croft}{1998}]%
        {Ponte98}
\bibfield{author}{\bibinfo{person}{Jay~M. Ponte} {and} \bibinfo{person}{W.
  Croft}.} \bibinfo{year}{1998}\natexlab{}.
\newblock \showarticletitle{A Language Modeling Approach to Information
  Retrieval}. In \bibinfo{booktitle}{{\em SIGIR}}. \bibinfo{pages}{275--281}.
\newblock


\bibitem[\protect\citeauthoryear{Radinsky, Svore, Dumais, Teevan, Bocharov, and
  Horvitz}{Radinsky et~al\mbox{.}}{2012}]%
        {Radinsky_etal_WWW2012}
\bibfield{author}{\bibinfo{person}{Kira Radinsky}, \bibinfo{person}{Krysta
  Svore}, \bibinfo{person}{Susan Dumais}, \bibinfo{person}{Jaime Teevan},
  \bibinfo{person}{Alex Bocharov}, {and} \bibinfo{person}{Eric Horvitz}.}
  \bibinfo{year}{2012}\natexlab{}.
\newblock \showarticletitle{Modeling and Predicting Behavioral Dynamics on the
  Web}. In \bibinfo{booktitle}{{\em WWW}}. \bibinfo{pages}{599--608}.
\newblock


\bibitem[\protect\citeauthoryear{Radinsky, Svore, Dumais, Shokouhi, Teevan,
  Bocharov, and Horvitz}{Radinsky et~al\mbox{.}}{2013}]%
        {radinsky2013behavioral}
\bibfield{author}{\bibinfo{person}{Kira Radinsky}, \bibinfo{person}{Krysta~M.
  Svore}, \bibinfo{person}{Susan~T. Dumais}, \bibinfo{person}{Milad Shokouhi},
  \bibinfo{person}{Jaime Teevan}, \bibinfo{person}{Alex Bocharov}, {and}
  \bibinfo{person}{Eric Horvitz}.} \bibinfo{year}{2013}\natexlab{}.
\newblock \showarticletitle{Behavioral Dynamics on the Web: Learning, Modeling,
  and Prediction}.
\newblock \bibinfo{journal}{{\em TOIS\/}} \bibinfo{volume}{31},
  \bibinfo{number}{3} (\bibinfo{year}{2013}), \bibinfo{pages}{Article 16}.
\newblock
\showISSN{1046-8188}


\bibitem[\protect\citeauthoryear{Rao, He, and Lin}{Rao et~al\mbox{.}}{2016}]%
        {rao2016noise}
\bibfield{author}{\bibinfo{person}{Jinfeng Rao}, \bibinfo{person}{Hua He},
  {and} \bibinfo{person}{Jimmy Lin}.} \bibinfo{year}{2016}\natexlab{}.
\newblock \showarticletitle{Noise-Contrastive Estimation for Answer Selection
  with Deep Neural Networks}. In \bibinfo{booktitle}{{\em CIKM}}.
  \bibinfo{pages}{1913--1916}.
\newblock


\bibitem[\protect\citeauthoryear{Rao, He, and Lin}{Rao et~al\mbox{.}}{2017}]%
        {rao2017cnnqa}
\bibfield{author}{\bibinfo{person}{Jinfeng Rao}, \bibinfo{person}{Hua He},
  {and} \bibinfo{person}{Jimmy Lin}.} \bibinfo{year}{2017}\natexlab{}.
\newblock \showarticletitle{Experiments with Convolutional Neural Network
  Models for Answer Selection}. In \bibinfo{booktitle}{{\em SIGIR}}.
\newblock


\bibitem[\protect\citeauthoryear{Rao and Lin}{Rao and Lin}{2016}]%
        {rao2016temporal}
\bibfield{author}{\bibinfo{person}{Jinfeng Rao} {and} \bibinfo{person}{Jimmy
  Lin}.} \bibinfo{year}{2016}\natexlab{}.
\newblock \showarticletitle{Temporal Query Expansion Using a Continuous Hidden
  Markov Model}. In \bibinfo{booktitle}{{\em ICITR}}.
  \bibinfo{pages}{295--298}.
\newblock


\bibitem[\protect\citeauthoryear{Rao, Lin, and Efron}{Rao
  et~al\mbox{.}}{2015}]%
        {rao2015reproducible}
\bibfield{author}{\bibinfo{person}{Jinfeng Rao}, \bibinfo{person}{Jimmy Lin},
  {and} \bibinfo{person}{Miles Efron}.} \bibinfo{year}{2015}\natexlab{}.
\newblock \showarticletitle{Reproducible Experiments on Lexical and Temporal
  Feedback for Tweet Search}.
\newblock In \bibinfo{booktitle}{{\em ECIR}}. \bibinfo{pages}{755--767}.
\newblock


\bibitem[\protect\citeauthoryear{Rao, Niu, and Lin}{Rao et~al\mbox{.}}{2016}]%
        {rao2016timeseries}
\bibfield{author}{\bibinfo{person}{Jinfeng Rao}, \bibinfo{person}{Xing Niu},
  {and} \bibinfo{person}{Jimmy Lin}.} \bibinfo{year}{2016}\natexlab{}.
\newblock \showarticletitle{Compressing and Decoding Term Statistics Time
  Series}. In \bibinfo{booktitle}{{\em ECIR}}. \bibinfo{pages}{675--681}.
\newblock


\bibitem[\protect\citeauthoryear{Rao, Ture, He, Jojic, and Lin}{Rao
  et~al\mbox{.}}{2017a}]%
        {rao2017talking}
\bibfield{author}{\bibinfo{person}{Jinfeng Rao}, \bibinfo{person}{Ferhan Ture},
  \bibinfo{person}{Hua He}, \bibinfo{person}{Oliver Jojic}, {and}
  \bibinfo{person}{Jimmy Lin}.} \bibinfo{year}{2017}\natexlab{a}.
\newblock \showarticletitle{Talking to Your TV: Context-Aware Voice Search with
  Hierarchical Recurrent Neural Networks}.
\newblock \bibinfo{journal}{{\em arXiv:1705.04892\/}}.
\newblock


\bibitem[\protect\citeauthoryear{Rao, Ture, Niu, and Lin}{Rao
  et~al\mbox{.}}{2017b}]%
        {rao2017termstatistics}
\bibfield{author}{\bibinfo{person}{Jinfeng Rao}, \bibinfo{person}{Ferhan Ture},
  \bibinfo{person}{Xing Niu}, {and} \bibinfo{person}{Jimmy Lin}.}
  \bibinfo{year}{2017}\natexlab{b}.
\newblock \showarticletitle{Mining Temporal Statistics of Query Terms for
  Searching Social Media Posts}. In \bibinfo{booktitle}{{\em ICTIR}}.
\newblock


\bibitem[\protect\citeauthoryear{Severyn and Moschitti}{Severyn and
  Moschitti}{2015}]%
        {severyn2015learning}
\bibfield{author}{\bibinfo{person}{Aliaksei Severyn} {and}
  \bibinfo{person}{Alessandro Moschitti}.} \bibinfo{year}{2015}\natexlab{}.
\newblock \showarticletitle{Learning to Rank Short Text Pairs with
  Convolutional Deep Neural Networks}. In \bibinfo{booktitle}{{\em SIGIR}}.
  \bibinfo{pages}{373--382}.
\newblock


\bibitem[\protect\citeauthoryear{Shen, He, Gao, Deng, and Mesnil}{Shen
  et~al\mbox{.}}{2014}]%
        {shen2014latent}
\bibfield{author}{\bibinfo{person}{Yelong Shen}, \bibinfo{person}{Xiaodong He},
  \bibinfo{person}{Jianfeng Gao}, \bibinfo{person}{Li Deng}, {and}
  \bibinfo{person}{{Gr\'{e}goire} Mesnil}.} \bibinfo{year}{2014}\natexlab{}.
\newblock \showarticletitle{A Latent Semantic Model with Convolutional-Pooling
  Structure for Information Retrieval}. In \bibinfo{booktitle}{{\em CIKM}}.
  \bibinfo{pages}{101--110}.
\newblock


\bibitem[\protect\citeauthoryear{Shokouhi and Radinsky}{Shokouhi and
  Radinsky}{2012}]%
        {Shokouhi_Radinsky_SIGIR2012}
\bibfield{author}{\bibinfo{person}{Milad Shokouhi} {and} \bibinfo{person}{Kira
  Radinsky}.} \bibinfo{year}{2012}\natexlab{}.
\newblock \showarticletitle{Time-Sensitive Query Auto-Completion}. In
  \bibinfo{booktitle}{{\em SIGIR}}. \bibinfo{pages}{601--610}.
\newblock


\bibitem[\protect\citeauthoryear{Smucker, Allan, and Carterette}{Smucker
  et~al\mbox{.}}{2007}]%
        {smucker2007comparison}
\bibfield{author}{\bibinfo{person}{Mark~D. Smucker}, \bibinfo{person}{James
  Allan}, {and} \bibinfo{person}{Ben Carterette}.}
  \bibinfo{year}{2007}\natexlab{}.
\newblock \showarticletitle{A Comparison of Statistical Significance Tests for
  Information Retrieval Evaluation}. In \bibinfo{booktitle}{{\em CIKM}}.
  \bibinfo{pages}{623--632}.
\newblock


\bibitem[\protect\citeauthoryear{Song, Elkahky, and He}{Song
  et~al\mbox{.}}{2016}]%
        {song2016multi}
\bibfield{author}{\bibinfo{person}{Yang Song}, \bibinfo{person}{Ali~Mamdouh
  Elkahky}, {and} \bibinfo{person}{Xiaodong He}.}
  \bibinfo{year}{2016}\natexlab{}.
\newblock \showarticletitle{Multi-Rate Deep Learning for Temporal
  Recommendation}. In \bibinfo{booktitle}{{\em SIGIR}}.
  \bibinfo{pages}{909--912}.
\newblock


\bibitem[\protect\citeauthoryear{Wang, Yu, Zhang, Gong, Xu, Wang, Zhang, and
  Zhang}{Wang et~al\mbox{.}}{2017}]%
        {wang2017irgan}
\bibfield{author}{\bibinfo{person}{Jun Wang}, \bibinfo{person}{Lantao Yu},
  \bibinfo{person}{Weinan Zhang}, \bibinfo{person}{Yu Gong},
  \bibinfo{person}{Yinghui Xu}, \bibinfo{person}{Benyou Wang},
  \bibinfo{person}{Peng Zhang}, {and} \bibinfo{person}{Dell Zhang}.}
  \bibinfo{year}{2017}\natexlab{}.
\newblock \showarticletitle{IRGAN: A Minimax Game for Unifying Generative and
  Discriminative Information Retrieval Models}.
\newblock \bibinfo{journal}{{\em arXiv:1705.10513\/}}.
\newblock


\bibitem[\protect\citeauthoryear{Zamani and Croft}{Zamani and Croft}{2017}]%
        {zamani2017relevance}
\bibfield{author}{\bibinfo{person}{Hamed Zamani} {and}
  \bibinfo{person}{W.~Bruce Croft}.} \bibinfo{year}{2017}\natexlab{}.
\newblock \showarticletitle{Relevance-based Word Embedding}. In
  \bibinfo{booktitle}{{\em SIGIR}}.
\newblock


\end{thebibliography}


\end{document}